\documentclass[prd,aps,twocolumn,nofootinbib]{revtex4-1}
\usepackage{amsfonts}
\usepackage{amsmath}
\usepackage{graphicx}
\usepackage{bm}

\usepackage{color}

\begin{document}

\title{EMRI data analysis with a phenomenological waveform}

\author{Yan Wang}
\email{yan.wang@aei.mpg.de} \affiliation{Albert Einstein Institute,
Callinstrasse 38, D-30167 Hanover, Germany\\and\\QUEST Centre for
Quantum Engineering and Space-Time Research, Leibniz University of
Hanover, Hanover, Germany}

\author{Yu Shang}
\email{Shang.Yu@aei.mpg.de} \affiliation{Albert Einstein Institute,
Am M\"{u}hlenberg 1 D-14476 Golm, Germany}

\author{Stanislav Babak}
\email{Stanislav.Babak@aei.mpg.de} \affiliation{Albert Einstein
Institute, Am M\"{u}hlenberg 1 D-14476 Golm, Germany}

\begin{abstract}
Extreme mass ratio inspirals (EMRIs) (capture and inspiral of a
compact stellar mass object into a Massive Black Hole (MBH)) are
among the most interesting objects for the gravitational wave
astronomy. It is a very challenging task to detect those sources
with the accurate estimation  parameters of binaries primarily due
to a large number of the secondary maxima on the likelihood surface.
Search algorithms based on the matched filtering require computation
of the gravitational waveform hundreds of thousands of times, which is
currently not feasible with the most accurate (faithful) models of
EMRIs. Here we propose to use a phenomenological template family
which covers a large range of EMRIs parameter space. We use these
phenomenological templates to detect the signal in the simulated
data and then, assuming a particular EMRI model, estimate the
physical parameters of the binary.  We have separated the detection
problem, which is done in a model-independent way, from the
parameter estimation. For the latter one, we need to adopt the model
for inspiral in order to map phenomenological parameters onto the
physical parameter characterizing EMRIs.

\end{abstract}
\maketitle
\section{Introduction}

  Stellar compact objects like a black hole, neutron star or white
dwarf in the cusp surrounding the massive black hole (MBH) in the
galactic nuclei could be deployed on a very eccentric orbit due to
N-body interaction. Such an object could either plunge (directly or
after few orbits) into MBH or form an EMRI: inspiralling compact
object on originally very eccentric orbit which shrinks and
circularizes due to loss of the energy and angular orbital momentum
through gravitational radiation. The compact object spends $\sim 10^6$ orbits in the very strong field of a MBH before it plunges, all this orbital
evolution will be encoded in the phase of emitted  gravitational
waves (GWs). Space based GW observatories, like LISA or similar
planned missions, will observe those sources few years before the
plunge. By fitting precisely the GW phase one can extract extremely
accurate parameters of a binary system \cite{Barack:2003fp} (like
mass and spin  of MBH $M, a$, mass of a small object $m$,
inclination of the orbital plane (to the spin of MBH), orbital
eccentricity and semi-latus rectum ($\iota_0, e_0, p_0$) at some
fiducial moment of time $t_0$, location of the source on the sky
($\theta, \phi$) and more).

   Precise tracking of the GW phase implies that we can also test the nature of the central massive object. The general
belief is that it should be a MBH with surrounding spacetime
described by a Kerr solution. The nature of the spacetime  affects
the orbital evolution of the compact object which in turn could be
extracted from the GW phase.  Kerr spacetime is described by only
two parameters: black hole's mass and spin, as stated by a
``no-hair'' theorem.    The spacetime could be decomposed in the
multipole moments of a central massive object, and, for Kerr BH, all
moments depend only on $M$ and $a$: $M_l + iS_l = (ia)^l M^{l+1}$
where $M_l$ and $S_l$ are mass and current moments. We could measure
three first moments (mass, spin and quadrupole
moment)\cite{Barack:2006pq}, and check the ``Kerrness'' of a
spacetime. In general, the deviations from Kerr could come in
several ways: (i) it is Kerr BH but there is an additional perturber
(gas disk, another MBH) (ii) it is not Kerr BH but some other object
satisfying GR (boson star, gravastar), (iii) there are deviations
from GR. For discussion on the topics we refer the reader to
\cite{Babak:2010ej, AmaroSeoane:2010zy, AmaroSeoane:2012je} and
references therein.

       Modeling  orbital evolution even within GR is not yet fully complete. Large mass ratio allows us to consider a small compact object
as a perturbation on the Kerr background spacetime, and treat the
problem perturbatively in orders of the mass ratio. In zero order
approximation the compact object moves on a geodesic orbit, however,
as soon as we assign the mass to it, it creates its own
gravitational field interacting with the background and this system
emits gravitational radiation.  The force resulting from the
interaction of the self field with the background is called self
force, and the motion of the compact object could be seen as the
forced geodesic motion.  Alternative interpretation is that the
motion is governed by a geodesic motion but in the perturbed
spacetime. Calculation of the self force is a complicated task which
is accomplished for the orbits around Schwarzschild BH only
\cite{Warburton:2011fk, Barack:2010tm}, the Kerr spacetime is
underway. There are also questions concerning the calculation of the
orbital evolution under the self force: the self force depends on
the past history of the compact object (which is usually assumed to
be a geodesic in the background spacetime). To compute the motion
under the self force one can use
 the osculating elements approach \cite{Gair:2010iv}, or self-consistent
approach of direct integration of the regularized equations
\cite{Diener:2011cc}. For more details on this subject we refer to
\cite{lrr-2011-7}.

 All in all, the modeling of the orbital evolution and the GW signal is a complex task which requires significant theoretical
and computational developments. The latter prevents us currently
from using the state-of-art GW models of EMRIs in our data analysis
explorations. In majority of the cases the phenomenological model
suggested in    \cite{Barack:2003fp}, so called ``analytic kludge''
(AK), is used. It is based on Post-Newtonian expressions and puts
together all relevant physics of EMRIs. However, this model has
restrictions in the number of harmonics and in their strength,  and
any search algorithm which relies on its specific harmonic content
will not work for a more realistic model of GW signal. The main
motivation of this work is to create the phenomenological search
template family which would fit a very large range of EMRI-like
signals. The typical EMRI signal consists of a set of harmonics of
three (slowly evolving) orbital frequencies, and we will use it as a
basis of our template. The phenomenological template consists of
$N_h$ harmonics with constant amplitude and slowly evolving phase
which we decompose in a Taylor series. Truncation of the Taylor
series and the assumption about constant amplitude set restrictions
on the duration over which the phenomenological template can fit an
EMRI signal.  The amplitude of EMRI's harmonics changes due to
shrinking of the orbit (overall amplitude increases),
circularization of the orbit (power is shifted to lower harmonics)
and slight change in the inclination of the orbit to the spin of
MBH.  Using more terms in the Taylor series helps to track phase of
the EMRI signal for longer time (which is more important than accurate
description of the amplitude). Finally, we decide on the number of
harmonics to use in the template (and their indices) based on the
analysis of the harmonic structure of the  Numerical Kludge (NK)
model \cite{Babak:2006uv} of EMRI in different parts of the
parameter space. The restriction that the phenomenological waveform
(PW) is valid only for a limited period of time  is very weak since
we can fit the signal piecewise, as long as the accumulated
signal-to-noise ratio (SNR) over that time is significant to claim
presence of the signal. In this work we consider only those parts of
the EMRI signal where the orbital frequencies are not decreasing
which is true over almost all time of the inspiral and breaks quite
close to the plunge. However, this is not really necessary since we
did not restrict the values of frequency derivatives to positive
values during the search.

The PW family is quite  generic and does not depend on the orbital
evolution, or, in other words, the orbital evolution of the binary
is encoded in the Taylor coefficients of phase of each harmonic.
This allows us to detect an EMRI signal in a model independent way.
Once the harmonics of the signal are recovered we can analyze them
using a specific EMRI model to recover physical parameters of the
system. It is at this point we need the orbital evolution with high
accuracy, which involves computation of the self-force and tests of
possible deviations from the ``Kerrness''.

 After constructing the phenomenological waveform we perform blind searches on the simulated data without noise (to avoid
stochastic errors in the parameter estimation) and with the noise.
We have used the NK waveform (as described in \cite{Babak:2006uv})
as a model of our signal and the orbital evolution according to
\cite{Gair:2005ih}. We have also used Markov chain Monte-Carlo
(MCMC) search  with phenomenological waveforms on the simulated
three month of data.  This search has provided us with multiple
local maxima in the likelihood which we gathered and analyzed in a
similar way as described in \cite {Babak:2009ua}. We associate local
maxima in the likelihood with partial detections of the signal and
construct the time-frequency map of the detected (patchy) harmonics
of the source. The next step is to assume the model for the orbital
evolution and, by matching the found time-frequency tracks to the
harmonics of the signal, estimate parameters of the binary system.
We have used the same model for the orbital evolution as in the
simulated data sets and recovered physical parameters with precision
better than few percent.

The paper is organized  as follows. In the next Section we will give
a brief overview of available models for GWs from EMRIs. In
Section~\ref{S:PW}, we introduce PW family in details.  We describe
MCMC search with PWs in Section~\ref{S:MCMC}. Analysis of MCMC
results and mapping to the physical parameters are done in the
Section~\ref{S:PSO}. Finally we conclude with a summary
Section~\ref{S:Summary}.

\section{Review of EMRI waveforms}
\label{S:REW}
    As was already mentioned in the Introduction, accurate computation of the GWs from EMRIs and the orbital evolution is
a complex and computationally intensive task.  The most promising
approach probably is the coupled integration of the compact object
dynamics and GW emission taken in    \cite{Diener:2011cc}.
Alternatively, one could have a separate evolution of the orbital
motion using self force computed across various geodesic orbits and
employ osculating elements approach \cite{Pound:2007th,
Gair:2010iv}. The waveform at infinity could be obtained from the
Teukolsky equations  \cite{Teukolsky:1973ha}  in time or in
frequency domain \cite{Martel:2003jj, Drasco:2005kz}.

The above methods are computationally expensive and several
approximations were suggested. Less accurate but still quite
reliable are Numerical Kludge (NK) waveforms: original NK
\cite{Babak:2006uv}
 and extended/improved NK called ``Chimera''  \cite{Sopuerta:2011te, Sopuerta:2012de}.
Those methods combine accurate prescription for the orbital evolution with approximate (Post-Newtonian)
waveform generation formalism.

The less precise model, which captures all relevant physics of EMRIs
(orbital precession, three orbital frequencies) was suggested in
\cite{Barack:2003fp}, so called Analytic Kludge  waveform. These
waveforms are very fast to generate, and even though they cannot be
used for searching for actual GW signals, they are used to develop
data analysis algorithms and to evaluate their performance
\cite{Barack:2003fp, Barack:2006pq, Babak:2009cj}.

In this work, we used NK waveform. In the original paper,
\cite{Babak:2006uv}, the waveform was generated in the time domain,
we have reimplemented it in the frequency domain following
suggestions of S. Drasco who did it first (private communications).
However, we still take into account the time dependence of only
three orbital constants under radiation reaction: energy, azimuthal
component of the orbital angular momentum and Carter constant. Under
the self force in osculating element approach we should evolve also
other three constants (defining initial position of the compact
object) due to conservative part of the self-force
\cite{Pound:2007th, Gair:2010iv}.  This does not affect our search
results, since PW is model independent, however, we have to use the
same model (as in the simulated data) for mapping the
phenomenological parameters onto the physical parameters of the
binary.  Mismatch in the models would result in the bias which we
want to avoid.

Finally we want to avoid using in this work the Analytic Kludge
model, because it predicts somewhat simplified (detectable) harmonic
content of the waveform.   The NK waveforms for generic orbits  were
compared against waveforms based on the Teukolsky equation and they
show quite good agreement.  We  believe that NK deviates from the
true EMRI signal in the phase but not so much in the number and
strength of harmonics. Therefore we use NK model as a representation
of the EMRI signal throughout this paper.

\section{EMRI phenomenological waveform family}
 \label{S:PW}
There are several algorithms which have been proven to be successful
in detecting EMRIs in the simulated LISA data \cite{Cornish:2008zd,
Babak:2009cj, Babak:2009ua}. However, those algorithms partially
utilize the features of AK waveform which was used in the simulation
of the data and in the data analysis. As explained in
Section~\ref{S:REW}, we want to avoid it by building a generic
phenomenological template family.

\subsection{Phenomenological waveform in the source frame}

The model we want to propose is based on the following assumptions
about GW signals from EMRIs:
\begin{enumerate}
\item The orbital motion can be effectively described by six slowly changing quantities.
Explicitly, three time-dependent initial phases are governed by the
conservative part of the self force; three fundamental
time-dependent frequencies are governed by the radiative part of the
self force.
\item The waveform is represented by harmonics of three
frequencies (phenomenologically, these frequencies are the summation
of the fundamental orbital frequencies and the evolution of the initial
phases) with slowly changing intrinsic amplitude:
\end{enumerate}
\begin{eqnarray}
h(t) &=& \sum_{l,m,n}{h_{lmn}(t)}
\nonumber \\&=& \mathrm{Re} \left(\sum_{l,m,n}{A_{lmn}(t)e^{i\Phi_{lmn}(t)}}\right) \nonumber \\
&=& \mathrm{Re} \left(\sum_{l,m,n}{A_{lmn}(t)e^{i(l \Phi_r + m
\Phi_{\theta} + n \Phi_{\varphi})}}\right),
\end{eqnarray}
where $\Phi_r,\Phi_{\theta},\Phi_{\varphi}$ are the phase evolutions
corresponding to the three fundamental motions.      Here we omitted
the tensorial spatial indices for simplicity.

The first assumption basically expresses that the motion is
described by a slow drift from one geodesic to another. The initial
phases correspond to the initial position of a compact object on a
given geodesic and the orbital frequencies are functions of the
energy, azimuthal component of the orbital momentum and Carter
constant. The slow drift ensures that phases $\Phi_{lmn}$ are slowly
varying functions of time.

\begin{figure}
\includegraphics[width=0.5\textwidth]{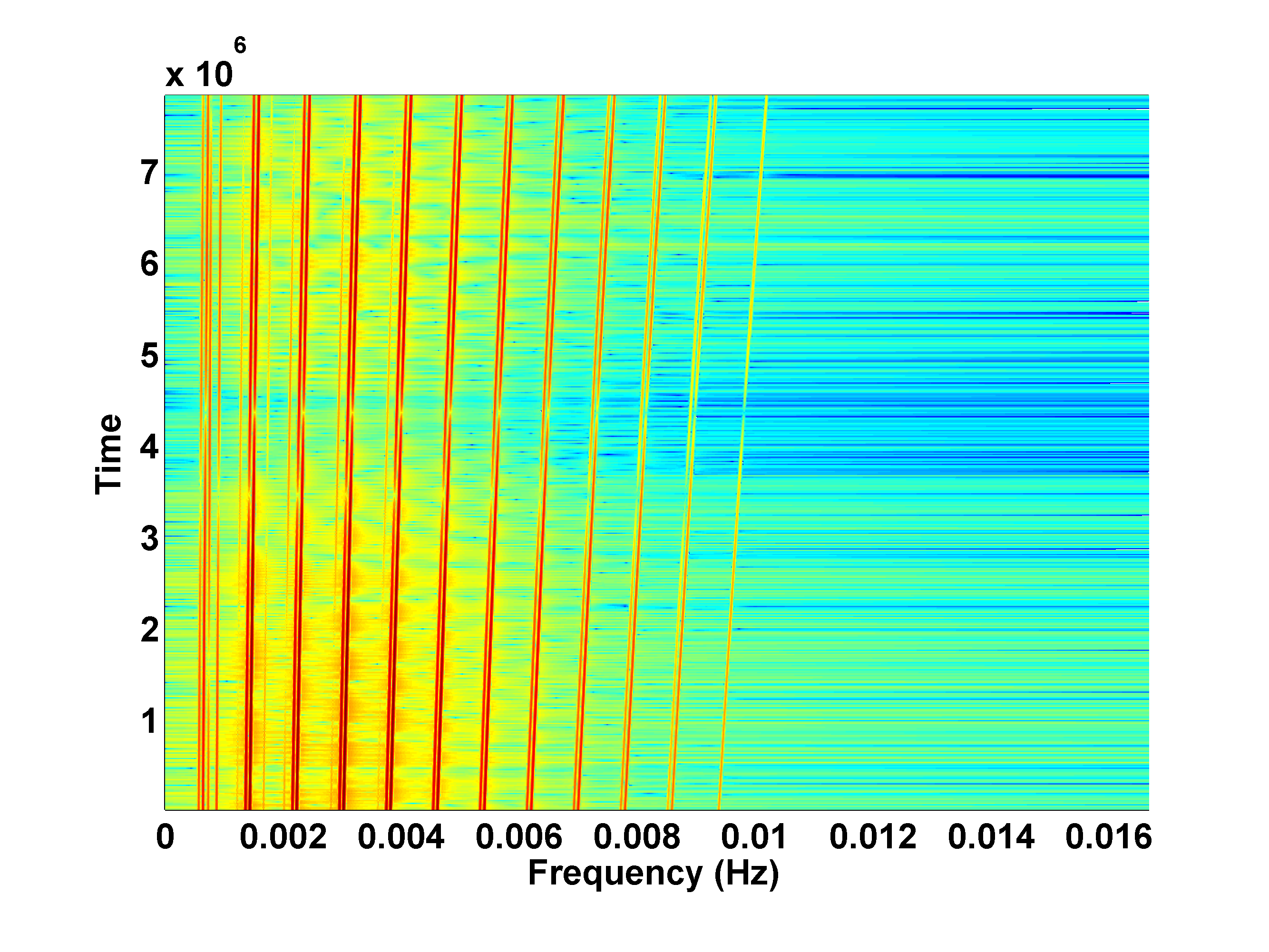}
\caption{ \label{fig:tf_plot1} The time-frequency plot of a typical
EMRI signal without noise. There are 30 dominant harmonics in
total.}
\end{figure}

Fig.~\ref{fig:tf_plot1} shows the time-frequency plot of a typical
EMRI signal.There are 30 clearly separated frequency tracks in the noiseless
plot, which display the dominant harmonics. It can also be seen that
the frequencies of harmonics are smooth and vary slowly.
It is generally true that both amplitude and the phase are slowly
varying functions of time, thus we can safely make the Taylor
expansion:
\begin{eqnarray}
\Phi_r(t) &=& \Phi_r(t_0) + \omega_r(t_0)(t-t_0)+\frac{1}{2}\dot{\omega}_r(t-t_0)^2+\ldots  \nonumber\\
&=&  \Phi_r(t_0) + 2\pi f_r(t_0)(t-t_0)+ \pi \dot{f}_r(t-t_0)^2+\ldots ,\nonumber \\
\\
\Phi_\theta(t) &=& \Phi_\theta (t_0) + \omega_\theta (t_0)(t-t_0)+\frac{1}{2}\dot{\omega}_\theta (t-t_0)^2+\ldots\nonumber \\
&=&  \Phi_\theta(t_0) + 2\pi f_\theta(t_0)(t-t_0)+ \pi \dot{f}_\theta(t-t_0)^2+\ldots ,\nonumber\\
\\
\Phi_{\varphi}(t) &=& \Phi_{\varphi}(t_0) + \omega_{\varphi}(t_0)(t-t_0)+\frac{1}{2}\dot{\omega}_{\varphi}(t-t_0)^2+\ldots \nonumber\\
&=&  \Phi_{\varphi}(t_0) + 2\pi f_{\varphi}(t_0)(t-t_0)+ \pi \dot{f}_{\varphi}(t-t_0)^2+\ldots ,\nonumber\\
\\
A_{lmn}(t) &=& A_{lmn}(t_0)+ \dot{A}_{lmn}(t_0)(t-t_0)+\ldots .
\end{eqnarray}
Since the amplitudes $A_{lmn}$ are even smoother than the phase over extended
period of time, and because the detection techniques are more sensitive to
mismatch in the phase than in the amplitude, we can neglect the time evolution
in the amplitudes  and treat all of them as constant. It is a very good assumption over
three months of the simulated data which we analyze in this paper.
 As for the phase expansion, we calculate the so-called fitting factor (FF)
for the different orders of polynomial approximations of the phase
to check the fidelity of the PW. Numerical results show that the
Taylor expansion for three months data, up to $\ddot{f}$ order,
gives the FF around $0.9$, and up to $\dddot{f}$ order the FF is
larger than $0.999$. So it is sufficient to expand the phase to
$\dddot{f}$ order. This is the phenomenological waveform family
which we propose to analyze an EMRI signal. To summarize, the
phenomenological waveform is a summation of individual harmonics
with constant (or linear) amplitudes and polynomial (in time)
phases.

\subsection{From the source frame to the LISA  frame}

First we will express the GW wavefrom in the solar system barycenter
frame and then translate it to the frame attached to LISA  (or a
LISA-like  space based observatory). In the source frame, an
arbitrary gravitational wave~(GW) signal in the TT gauge can be
written in the following form:
\begin{equation}
\mathbf{h}(t)=h^S_+(t)\mathbf{e}_+ + h^S_\times(t)\mathbf{e}_\times
\end{equation}
where the superscript 'S' denotes the source frame. Since the LISA
constellation is orbiting the sun, it is convenient
to express the GW signal in the solar system barycenter (SSB) frame.
\begin{eqnarray}
\mathbf{h}(t)&=&h_+(t)\bm{\epsilon}_+ +
h_\times(t)\bm{\epsilon}_\times\\
\bm{\epsilon}_+&=&\hat{\theta}^S\otimes\hat{\theta}^S-\hat{\phi}^S\otimes\hat{\phi}^S\\
\bm{\epsilon}_\times&=&\hat{\theta}^S\otimes\hat{\phi}^S+\hat{\phi}^S\otimes\hat{\theta}^S
\end{eqnarray}
where $(\theta^S,\phi^S)$ denotes the direction of the GW source in
the SSB frame, $\hat{\theta}^S,\hat{\phi}^S$ are the unit vectors
along longitudinal and latitudinal directions.
The principal polarization vectors attached to the solar system barycenter
frame,   $\hat{\theta}^S,\hat{\phi}^S$  are connected to the
principal polarization vectors in the source frame via rotation angle
$\psi$ (since they lie in the same plane orthogonal to the GW propagation
direction). The polarization
components $h_+$ and $h_\times$ are transformed under this rotation according to
\begin{eqnarray}
h_+ &=& h^S_+ \cos(2\psi)+h^S_\times \sin(2\psi)\\
h_\times &=& -h^S_+ \sin(2\psi)+h^S_\times \cos(2\psi).
\end{eqnarray}

Now we will add LISA response.
LISA measures the Doppler shift of the inter-spacecraft lasers
induced by a gravitational wave. The single-link full response to this
frequency shift can be derived with the help of three Killing
vectors~\cite{Estabrook:1975}.
However, this single-link signal is orders of magnitude smaller than
the dominating laser frequency noise. Thus, we need to use the
so-called Time-Delay-Interferometry (TDI) variables~\cite{Armstrong99},
which cancel the laser noise through the recombination of the
artificially delayed single-link signals. In the low frequency limit, the two orthogonal TDI
 (noise independent) variables of  Michelson type can be expressed
as~\cite{Cutler:1997ta, Rubbo:2003ap}
\begin{eqnarray}
h_{I}(t)&=&[\delta L_1(t)-\delta L_2(t)]/L \nonumber \\
&=&\mathbf{h}(\zeta):\bm{D}_I \\
h_{II}(t)&=&\frac{1}{\sqrt{3}}[\delta L_1(t)+\delta L_2(t)-2\delta L_3(t)]/L \nonumber \\
&=&\mathbf{h}(\zeta):\bm{D}_{II}
\end{eqnarray}
where $L$ stands for the average arm length. The retarded time
$\zeta(t)=t-\hat{k}\cdot\mathbf{x}/c$ defines the wavefront, where
$\hat{k}$ is the GW propagation direction. The two detector tensors
are defined as $\bm{D}_I\equiv\frac{1}{2}(\hat{n}_1\otimes
\hat{n}_1-\hat{n}_2\otimes
\hat{n}_2),\bm{D}_{II}\equiv\frac{1}{2\sqrt{3}}(\hat{n}_1\otimes
\hat{n}_1+\hat{n}_2\otimes \hat{n}_2-2\hat{n}_3\otimes \hat{n}_3)$,
where $\hat{n}_1, \hat{n}_2, \hat{n}_3$ denote the unit vectors
along each arm of LISA. Here we assume LISA-like setup which has six
links (three arms).  Even though the EMRI signal could reach quite
high frequencies and require full response, we adopt the
low-frequency approximation for our exercises. This does not
restrict ability of our analysis as long as the simulated signal and
the search template use the same response function.

\subsection{Data analysis with phenomenological waveform.}
\label{S:DAFW}

We start with a brief overview of our notations and basics of data analysis.
We denote the Fourier transform of a time series $a(t)$ by
$\tilde{a}(f)$ and adopt the following convention
\begin{eqnarray}
\tilde{a}(f)=\int a(t)e^{i2\pi f t}dt.
\end{eqnarray}
We assume that the detector is characterized by a Gaussian,
stationary noise $n(t)$ and its two-sided noise power spectral
density is defined as
$\overline{\tilde{n}^*(f')\tilde{n}(f)}=S_n(f)\delta(f-f')$, where
the over bar denotes the ensemble average. With this power spectral
density, it is conventional to define an inner product of two time
series $a(t),b(t)$ as follows
\begin{eqnarray}
<a|b>=\int_{-\infty}^{\infty}\frac{\tilde{a}^*(f)\tilde{b}(f)}{S_n(f)}df.
\end{eqnarray}
The signal-to-noise ratio is defined as
\begin{eqnarray}
SNR^2\equiv<h|h>=\int_{-\infty}^{\infty}\frac{|\tilde{h}(f)|^2}{S_n(f)}df.
\end{eqnarray}
where $h$ is the GW signal. Let us denote the probability of a
gravitational wave signal $h(\bm{\theta})$ being present in the data
$s(t)$ by $P(s|h(\bm{\theta}))$, where $\bm{\theta}$ is the set of
parameters that characterizes the gravitational wave signal.
Similarly, the probability of no gravitational wave signal
present in the data $s$ is denoted by $P(s|0)$. Likelihood ratio
$\Lambda(\bm{\theta})$ is the ratio between these two probabilities
\begin{eqnarray}\label{likelihood}
\Lambda(\bm{\theta})&=&\frac{P(s|h(\bm{\theta}))}{P(s|0)}\nonumber\\
&=&e^{<s|h(\bm{\theta})>-\frac{1}{2}<h(\bm{\theta})|h(\bm{\theta})>}.
\end{eqnarray}
 It is conventional to consider rather logarithm of the likelihood ratio
as a detection statistic:
$L(\bm{\theta})=\log\Lambda(\bm{\theta})=<s|h(\bm{\theta})>-\frac{1}{2}<h(\bm{\theta})|h(\bm{\theta})>$.
This is the quantity we want to maximize over the parameter set
$\bm{\theta}$.

\begin{figure}
\includegraphics[width=0.5\textwidth]{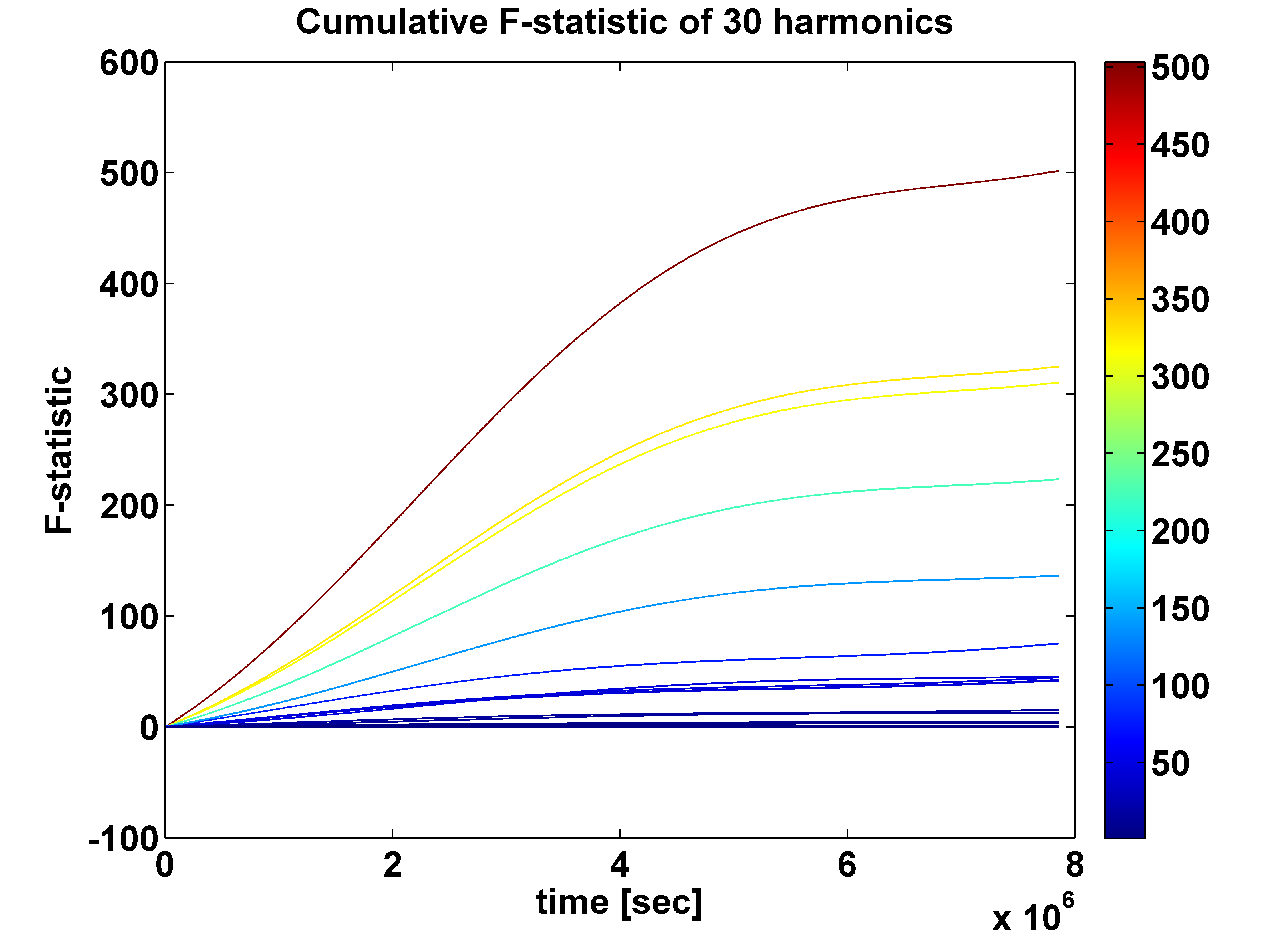}
\caption{ \label{fig:cumuF1} The cumulative F-statistic of 30
dominant harmonics with true parameters without noise. Since there
is no noise, the F-statistic is not normalized.}
\end{figure}

\begin{figure}
\includegraphics[width=0.5\textwidth]{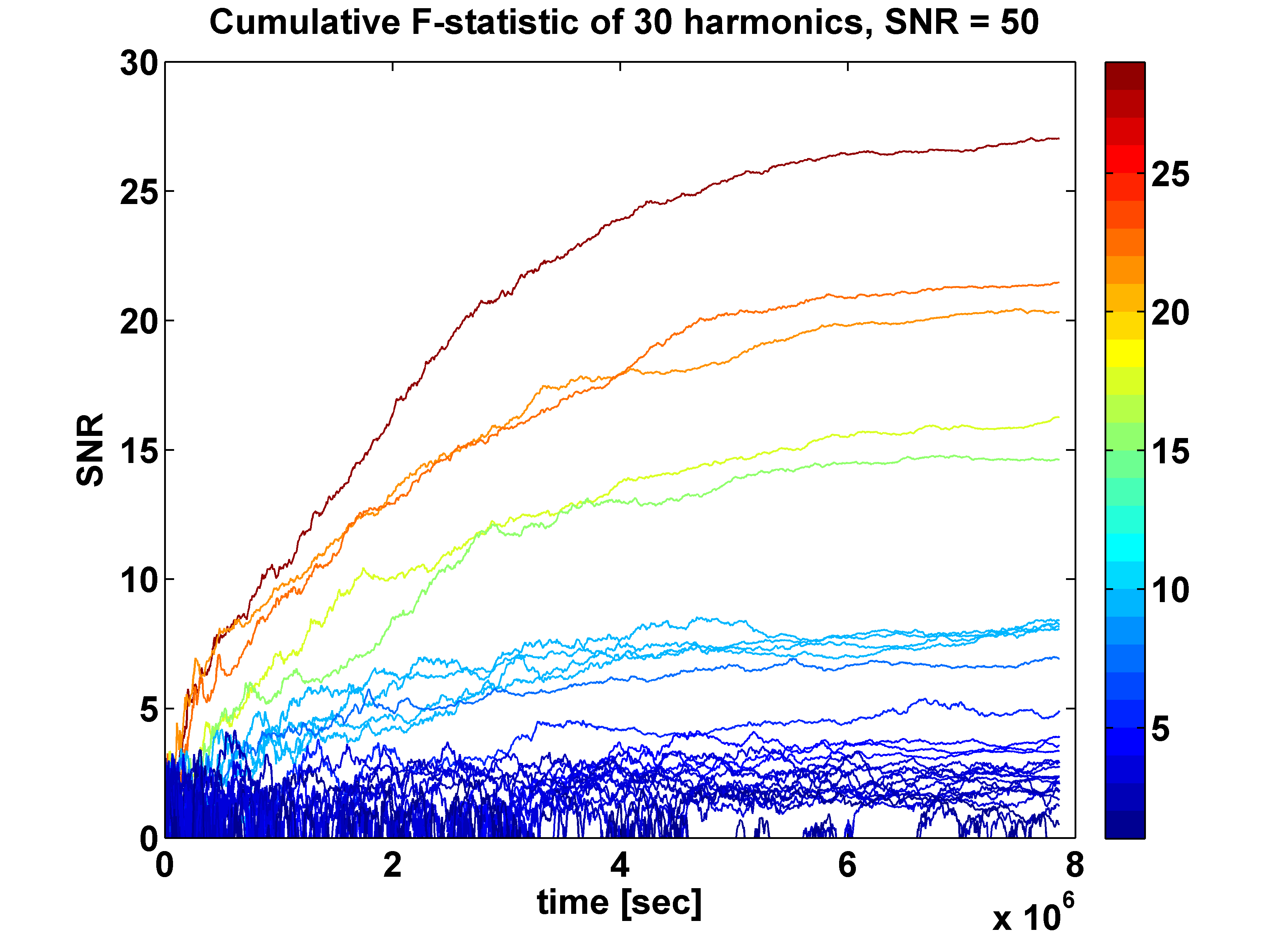}
\caption{ \label{fig:cumuFsnr50} The cumulative F-statistic of 30
dominant harmonics with true parameters and detector noise. Note
that the F-statistic is converted to $SNR$ in the figure. The strong
harmonics are cumulating gradually with local spikes. The low-SNR
harmonics behave similar to noise, hence made undetectable.}
\end{figure}

The likelihood ratio could be further simplified if we use PW.
A single harmonic with polynomial phase up to $\dddot{f}$ order in
the source frame takes the following form
\begin{eqnarray}
\mathbf{h}(t)&=&A_+\cos(\Phi(t)+\Phi_0)\mathbf{e}_+ + A_\times
\sin(\Phi(t)+\Phi_0)\mathbf{e}_\times ,\nonumber\\
\\
\Phi(t)&=& 2\pi f(t-t_0) + \pi \dot{f}(t-t_0)^2 +\nonumber\\
&&\frac{\pi}{3}\ddot{f}(t-t_0)^3 + \frac{\pi}{12}\dddot{f}(t-t_0)^4,
\end{eqnarray}
where we have omitted harmonic indices  $l,m,n$.
After simple algebra, LISA's response to this single harmonic GW
signal without noise can be put in a simple form
\begin{eqnarray}
  \label{E:Dec}
h_I(t) = A^\mu h_\mu^I(t), \;\;\;\;
h_{II}(t) = A^\mu h_\mu^{II}(t)
\end{eqnarray}
where we follow summation convention over repeated indices,
and $\mu=1,2,3,4$. The four amplitude parameters $A^\mu$ depend only
on $(A_+,A_\times,\Phi_0,\psi)$,which are usually called extrinsic
parameters, while $h_\mu^I(t),h_\mu^{II}(t)$ are functions of
$(\theta^S,\phi^S,f,\dot{f},\ddot{f},\dddot{f})$, which are usually
called intrinsic parameters. From now on, we denote the intrinsic
parameters by $\bm{\theta}$. The extrinsic parameters  (being constants in our
approximation) can be maximized over analytically~\cite{Jaranowski:1998qm, Prix:2007zh},
which we will show explicitly below. We denote the measured data with noise
corresponding to $h_I(t),h_{II}(t)$ by $s_I(t),s_{II}(t)$. Since the
joint probability of a GW signal present in both $s_I$ and $s_{II}$
is just the product of the individual probabilities, the joint log
likelihood is just the summation of the individual log likelihoods
\begin{eqnarray}
L(\bm{\theta},A^\mu)&=&<s_I|h_I(\bm{\theta})>-\frac{1}{2}<h_I(\bm{\theta})|h_I(\bm{\theta})>\nonumber\\
&&+<s_{II}|h_{II}(\bm{\theta})>-\frac{1}{2}<h_{II}(\bm{\theta})|h_{II}(\bm{\theta})>.
\end{eqnarray}
Substituting (\ref{E:Dec}) into this expression  we arrive at
  \begin{eqnarray}
L(\bm{\theta},A^\mu)&=&A^\mu s^I_\mu(\bm{\theta})-\frac{1}{2}A^\mu
M^I_{\mu\nu}(\bm{\theta})A^\nu \nonumber\\
&+& A^\mu s^{II}_\mu(\bm{\theta})-\frac{1}{2}A^\mu
M^{II}_{\mu\nu}(\bm{\theta})A^\nu,
\end{eqnarray}
where we have used the following conventions:
$s^I_\mu=<s_I|h^I_\mu>$, $s^{II}_\mu=<s_{II}|h^{II}_\mu>$,
$M_{\mu\nu}^I=<h^I_\mu|h^I_\nu>$, $M_{\mu\nu}^{II}=<h^{II}_\mu|h^{II}_\nu>$.
We can maximize the log-likelihood over extrinsic parameters by solving
\begin{eqnarray}
\frac{\partial L(\bm{\theta},A^\mu)}{\partial
A^\mu}=(s^I_\mu+s^{II}_\mu)-(M^I_{\mu\nu}+M^{II}_{\mu\nu})A^\nu=0,
\end{eqnarray}
which is  straightforward to find
$A^\mu=[(M^I+M^{II})^{-1}]^{\mu\nu}(s^I_\nu+s^{II}_\nu)$.
 The log-likelihood maximized over the extrinsic parameters is called F-statistic:
\begin{eqnarray}\label{F-statistic}
F(\bm{\theta})&\equiv& \max_{A^\mu} L(\bm{\theta},A^\mu)\nonumber\\
&=&\frac{1}{2}(s^I_\mu+s^{II}_\mu)[(M^I+M^{II})^{-1}]^{\mu\nu}(s^I_\nu+s^{II}_\nu).
\end{eqnarray}
Its expectation   value is connected to the SNR in the following way
\begin{eqnarray}
E[F(\bm{\theta})]=\frac{1}{2}SNR^2+2.
\end{eqnarray}
Since $h(\bm{\theta})$ is narrow band signal, the inner product can
be written in the following form
\begin{eqnarray}
<a|b>&=&\int_{-\infty}^{\infty}\frac{\tilde{a}^*(f)\tilde{b}(f)}{S_n(f_0)}df\nonumber\\
&=& \frac{1}{S_n(f_0)} \int_{0}^{T} a(t)b(t)dt
\end{eqnarray}
where $T$ is the observation time, $f_0$ is the middle frequency of
$h(\bm{\theta})$. The inner product is a function of $T$, and so is
F-statistic. By varying $T$ from $0$ to the total observation time,
we define a {\em cumulative F-statistic} $F(T,\bm{\theta})$. The
cumulative F-statistic for 30 dominant harmonics without detector
noise is plotted in Fig.~\ref{fig:cumuF1}. The case with the
simulated detector noise is shown in Fig.~\ref{fig:cumuFsnr50}, the
total SNR of the signal in this case is $SNR=50$. Those are two data
sets which we will analyze in the next section.

 The cumulative F-statistic
provides much more information than F-statistic. Actually, if
$\bm{\theta}_*$ is the true parameter set of the signal, one can
argue that
\begin{eqnarray}
E\left[\frac{\partial F(T,\bm{\theta}_*)}{\partial T}\right]\propto
h^2(T) \xi^2(T),
\end{eqnarray}
where $\xi(T)=\sqrt{\xi_+^2(T)+\xi_\times^2(T)}$ is the geometrical
mean of the antenna pattern functions for two polarizations. When
there is no detector noise,  $\frac{\partial
F(T,\bm{\theta}_*)}{\partial T}=E\left[\frac{\partial
F(T,\bm{\theta}_*)}{\partial T}\right]$ is nonnegative. Thus,
$E[F(T,\bm{\theta}_*)]$ is always increasing over the entire time
span when the GW signal is present, as can be seen in
Fig.~\ref{fig:cumuF1}. It is not necessarily so in presence of the
noise and during analysis of the data. There are three types of
oscillations on the cumulative F-statistic curve $F(T,\bm{\theta})$.
(i). The (non-negative) oscillation due to the oscillatory nature of
the gravitational wave signal. It is at twice the GW frequency,
which makes it hard to see in Fig.~\ref{fig:cumuF1}. (ii). In
reality, we do not know the exact true parameters of the GW signal.
That means, in most cases, the parameter set $\bm{\theta}$ we try
differs from the true parameter set $\bm{\theta}_*$. This introduce
beat-notes to $F(T,\bm{\theta})$. This kind of oscillation happens
at beat-note frequency, which is much lower than the GW frequency
itself. (iii). The third type of oscillation is due the noise. The
presence of the noise makes the cumulative F-statistic uneven, see
Fig.~\ref{fig:cumuFsnr50}. Comparing to the former two types, this
kind of oscillation is irregular; it oscillates at all frequencies
and could cause temporary (for a short time) decrease in the
cumulative F-statistic.

We have found that over three months of simulated data we can
consider all harmonics as being completely independent with
virtually zero overlap  between them, $<h_{lmn} | h_{l'm'n'}> =
\delta_{l l'} \delta_{m m'} \delta_{n n'} $.
 The total F-statistic is therefore a sum of F-statistics from each
harmonic. In the next section we describe the search where we use
Eq. (\ref{F-statistic}) as a detection statistic, and we will use
cumulative F-statistic later on to analyze our findings.


\section{Search with  the phenomenological waveform}
 \label{S:MCMC}

In this section, we use the PW as described above together
with the introduced detection statistic.
  We will use two 3 month worth simulated data sets: with and without noise.
We use the same GW signal (based on NK model) in both cases.
The total SNR of the source in the noisy case is 50.
 We have taken the following parameters for the EMRI :
the mass of the MBH $M=10^6M_\odot$, the mass of the compact  object
(stellar mass BH)  $m=10M_\odot$, the initial orbital eccentricity
$e=0.4$, the semi-latus rectum $p=8M$, the inclination angle
$\iota=\pi/9$, the spin of the MBH $a=0.9M$, the sky position of the
source $(\theta^S,\phi^S)=(\pi/4,\pi/4)$, the polarization angle
$\psi=0$. In our analysis we assume that the sky location is known.
Our primary goal here is to recover the intrinsic parameters of the
source.

The noiseless case is used to avoid any possible bias in the final
result due to stochastic nature of the noise, and assess possible
restrictions of our search technique and PW family. Next, we apply
the same search method to the same GW signal buried in the noise,
which would justify its effectiveness in practice.

Here, we describe the search for individual harmonics with {\em
Markov chain Monte Carlo} (MCMC) method. For completeness and future
references we give a brief introduction to MCMC. Like a standard
Monte Carlo integration, MCMC is a random sampling method. It is
nothing but Monte Carlo integration with a Markov chain. By properly
constructing a Markov chain, MCMC can draw samples from the
searching parameter space more efficiently. Among all the
methodologies of constructing a Markov chain, the
Metropolis-Hastings scheme would be the most general one. The main
idea of Metropolis-Hastings algorithm is to cleverly construct a
Markov chain that satisfy the detailed balance equation, so that the
sampling distribution will converge to the likelihood surface we
want to estimate. If the shape of the likelihood surface is known,
the parameter set that corresponds to the maximum likelihood is
automatically known. Thus, MCMC is also widely used as a stochastic
optimization tool in GW data analysis (we refer the reader to a very
nice overview and discussion on Bayesian methods in
\cite{Littenberg:2009bm}, see also references therein).

 If the likelihood surface is multimodal(i.e. contains large number of separated
local maxima) then simple version of the MCMC finds a maximum and
does not move off it to explore larger parameter space. Many ways
around this problem were suggested but we will not use any of them
here (besides simulated annealing which we will discuss a later). As
we will see, a simple Metropolis-Hastings algorithm is sufficient.
The likelihood surface of an EMRI signal is very rich in ``wall''
and ``needle'' like structures, which make it very hard to find a
global maximum. We are interested in detecting as many local maxima
as possible. Therefore we run multiple independent chains and
harvest the results after they converge to various maxima of the
likelihood surface. If we are lucky, the global maximum could be
among multiple maxima we have found.

To understand the Metropolis-Hastings algorithm, first consider a
stochastic process denoted by $\{\bm{\theta}_k|k=0,1,2...\}$ which
belongs to the parameter space $\mathcal{B}$ in $\mathbb{R}^n$. Here
we defined $\bm{\theta}_k$ as
 a set of parameters at step $k$, which can also be viewed as a point in the parameter space
$\mathcal{B}$. If there exists a transition probability
$P(\bm{\theta}_{k+1}|\bm{\theta}_k)$ depending only on the current
point $\bm{\theta}_k$ for the stochastic process to be in state
$\bm{\theta}_{k+1}$, we call this stochastic process
$\{\bm{\theta}_{k}|k=0,1,2...\}$ a {\em Markov chain} with a
transition probability $P(\bm{\theta}_{k+1}|\bm{\theta}_k)$. In a
Bayesian viewpoint, we can take this transition probability as
conditional probability and immediately see that

\begin{equation}
\int_\mathcal{B}P(\bm{\theta}_{k+1}|\bm{\theta}_k)d\bm{\theta}_{k+1}=1\;.
\end{equation}

A Markov chain satisfying the {\em detailed balance equation}

\begin{equation}
\Lambda(\bm{\theta}_k)P(\bm{\theta}_{k+1}|\bm{\theta}_k)=\Lambda(\bm{\theta}_{k+1})P(\bm{\theta}_k|\bm{\theta}_{k+1})
\end{equation}
will (up to some relatively weak conditions) be equivalent to the
samples from the distribution $\Lambda(\bm{\theta})$ after a certain
initial period (often called {\em burn-in} stage). We can easily
estimate the distribution $\Lambda(\bm{\theta})$ with the Markov
chain samples and hence the most probable parameter set
$\hat{\bm{\theta}}$ for given observed data $\bf {s}$, where

\begin{equation}
\Lambda(\hat{\bm{\theta}}|\bf{s})=\max_{\bm{\theta}}\Lambda(\bm{\theta}|\bf{s})
\end{equation}
is usually called the maximum likelihood estimator.\\
By virtue of Metropolis-Hastings algorithm, we can construct a
Markov chain that satisfies the detailed balance equation and make
use of the corresponding property to estimate our template
parameters $\bm{\theta}$. To do this, we randomly choose a parameter
set $\bm{\theta}_0$ in the parameter space as the starting point.
Then one can pick a {\em proposal distribution}
$q(\bm{\theta}_{k+1}|\bm{\theta}_k)$ (as long as there is no
forbidden region in the prescribed parameter space to the point
$\bm{\theta}_{k+1}$) and sample a candidate point
$\bm{\theta}_{k+1}$ from this distribution. Then we calculate the
{\em acceptance probability} defined by the following formula

\begin{equation}\label{alpha}
\alpha(\bm{\theta}_k,\bm{\theta}_{k+1})=\min\left(1,\frac{\Lambda(\bm{\theta}_{k+1})
q(\bm{\theta}_k|\bm{\theta}_{k+1})}{\Lambda(\bm{\theta}_k)
q(\bm{\theta}_{k+1}|\bm{\theta}_k)}\right)\;.
\end{equation}
By accepting the point $\bm{\theta}_{k+1}$ according to the above
probability, we have, in fact, succeeded to construct a transition
probability,

\begin{equation}
P(\bm{\theta}_{k+1}|\bm{\theta}_k)=q(\bm{\theta}_{k+1}|\bm{\theta}_k)\alpha(\bm{\theta}_k,\bm{\theta}_{k+1})\;.
\end{equation}

It is easy to see that the Markov chain generated by the above
transition probability satisfies the detailed balance equation:

\begin{eqnarray}
\Lambda(\bm{\theta}_k)P(\bm{\theta}_{k+1}|\bm{\theta}_k) & = &
\min\left(\Lambda(\bm{\theta}_k)q(\bm{\theta}_{k+1}|\bm{\theta}_k),
\Lambda(\bm{\theta}_{k+1})q(\bm{\theta}_k|\bm{\theta}_{k+1})\right)\nonumber\\
& = &
\min\left(\Lambda(\bm{\theta}_{k+1})q(\bm{\theta}_k|\bm{\theta}_{k+1}),
\Lambda(\bm{\theta}_k)q(\bm{\theta}_{k+1}|\bm{\theta}_k)\right)\nonumber\\
& = & \Lambda(\bm{\theta}_{k+1})P(\bm{\theta}_k|\bm{\theta}_{k+1}).
\end{eqnarray}

Thus, such a Markov chain will eventually serve as a succession of
samples from $\Lambda(\bm{\theta})$. The best performance is
achieved  if the proposal probability
$q(\bm{\theta}_{k+1}|\bm{\theta}_k)$ resembles the target
distribution $\Lambda(\bm{\theta})$ over the entire parameter space.
Without prior knowledge about the kind of probability distribution
around the true parameter location, it is natural to choose it as a
multivariate normal distribution centered at the present point
$\bm{\theta}_k$ with covariance matrix $\mathcal{C}$,

\begin{equation}
q(\bm{\theta}_{k+1}|\bm{\theta}_k)=\frac{1}{\sqrt{(2\pi)^N
\mathrm{det}[\mathcal{C}]}}
\exp{\left[-\frac{1}{2}(\bm{\theta}_{k+1}-\bm{\theta}_k)^\textsf{T}\mathcal{C}^{-1}(\bm{\theta}_{k+1}-\bm{\theta}_k)\right]}\;,
\label{Eq:q}
\end{equation}
where $N$ denotes the dimension of the parameter space and
$\mathrm{det}[\mathcal{C}]$ the determinant of the covariance matrix
$\mathcal{C}$. The likelihood surface has usually multimodal
(multiple local maxima) structure, and, therefore, a single
multivariate normal distribution cannot describe the probability
density over the entire template space but only a very small region
around the local maximum. Since the probability distribution at the local maximum is usually
very sharp, a Markov chain easily gets trapped there for many steps.
To avoid insignificant maxima we use  the so-called {\em annealing}
scheme, originating from simulated annealing. We adopt two types of
annealing techniques: (i). we introduce a temperature
$\mathcal{T}_1$ to the acceptance rate $\alpha$ [equation
(\ref{alpha})] so as to have a larger possibility to accept the
proposal point in the beginning. By combining equations
(\ref{likelihood}), (\ref{F-statistic}), (\ref{alpha}),
(\ref{Eq:q}), the acceptance probability is now written as
\begin{eqnarray}\label{alpha_new}
\alpha(\bm{\theta}_k,\bm{\theta}_{k+1})=\min\left(1,e^{[F(\bm{\theta}_{k+1})-F(\bm{\theta}_k)]/\mathcal{T}_1}\right)\;.
\end{eqnarray}
where the temperature $\mathcal{T}_1 = \mathcal{T}_1(k) $ is a
function of the step index $k$, it starts from some relatively large
number and gradually decays to unity. (ii). We introduce a second
temperature $\mathcal{T}_2$ to the proposal distribution
$q(\bm{\theta}_{k+1}|\bm{\theta}_k)$. The covariance matrix
$\mathcal{C}$ is replaced by $\mathcal{C}\times \mathcal{T}_2$. Same
as $\mathcal{T}_1$, $\mathcal{T}_2$ is also a function of the step
index $k$, decaying gradually to unity. Hence, the chain take large
steps in the beginning and explores large volume in the parameter
space. Explicitly, we choose $\mathcal{T}_1$ and $\mathcal{T}_2$
both as a linear function
of $k$ with negative slope.\\
Let us summarize the algorithm:
\begin{enumerate}
\item $k=0$. Choose a random parameter set $\bm{\theta}_0$ as the starting point and
calculate the F-statistic $F(\bm{\theta}_0)$.
\item $k\rightarrow k+1$. Calculate the temperature $\mathcal{T}_1(k), \mathcal{T}_2(k)$.
\item Generate the next candidate parameter set $\bm{\theta}_{c}$
from the proposal distribution with modified covariance
$\mathcal{C}\times \mathcal{T}_2$.
\item Calculate the F-statistic of the new parameter set
$F(\bm{\theta}_{c})$.
\item Calculate the acceptance probability $\alpha(\bm{\theta}_k,\bm{\theta}_{c})=\min\left(1,e^{[F(\bm{\theta}_{c})-F(\bm{\theta}_k)]/\mathcal{T}_1}\right)$.
\item Draw a random number $u$ from unity distribution
$\mathcal{U}(0,1)$. If $u<\alpha$, accept the candidate parameter
set $\bm{\theta}_{k+1}=\bm{\theta}_{c}$, else, stay at the current
point $\bm{\theta}_{k+1}=\bm{\theta}_{k}$.
\end{enumerate}
 In the search we have used a diagonal form of the covariance matrix in the gaussian
proposal distribution (\ref{Eq:q}), with the following elements:
  $\mathcal{C} = [\mathrm{diag}(10^{-4},\, 10^{-12},\, 10^{-20},\, 10^{-28})]^2$ corresponding  to the parameter set
$\{f, \dot{f}, \ddot{f}, \dddot{f}\}$.    And $\mathcal{T}_2$ used to scale the covariance matrix
 decays   linearly with the number of members in the chain from 1 to $5 \times 10^{-4}$.
We have found that the use of the actual Fisher information matrix
as $\mathcal{C}$ did not improve significantly the search results.
We run about 50 chains on both noiseless data and noisy data. All
the parameter sets that generate an SNR larger than a certain
threshold (we have used  $SNR>4.5$) are recorded. Notice that there
are possibly many such qualified parameter sets in a single chain.
Thus, we have hundreds to thousands of qualified parameter sets or
local maxima. These local maxima contain information about the
signal. We will analyze these local maxima in the next section.

\section{Analysis of  the search results and mapping to the physical parameters}
 \label{S:PSO}

In this section we will explain how we use the results of MCMC
search described in the previous section and reconstruct harmonics
of the GW signal. Furthermore, we use the model of EMRI (NK) to
estimate the physical parameters of the system.

\subsection{Clustering algorithms}

In this subsection we extract information from the local maxima
detected by MCMC search. We first focus on the noiseless data to
explain the algorithm, then modify it a bit and apply it to the
noisy data. Since this work is the first of a series of papers, the
main task here is to establish the framework and justify the method.
Hence, as mentioned above,  we have  assumed that the sky position
of the  source is known and concentrate on the intrinsic parameters
only. This will save us some time, yet maintain all the main
features of the problem. As a result, each local maximum is
characterized  only  by  the frequency and its derivatives
$(f,\dot{f},\ddot{f},\dddot{f})$.

\begin{figure}
\includegraphics[width=0.5\textwidth]{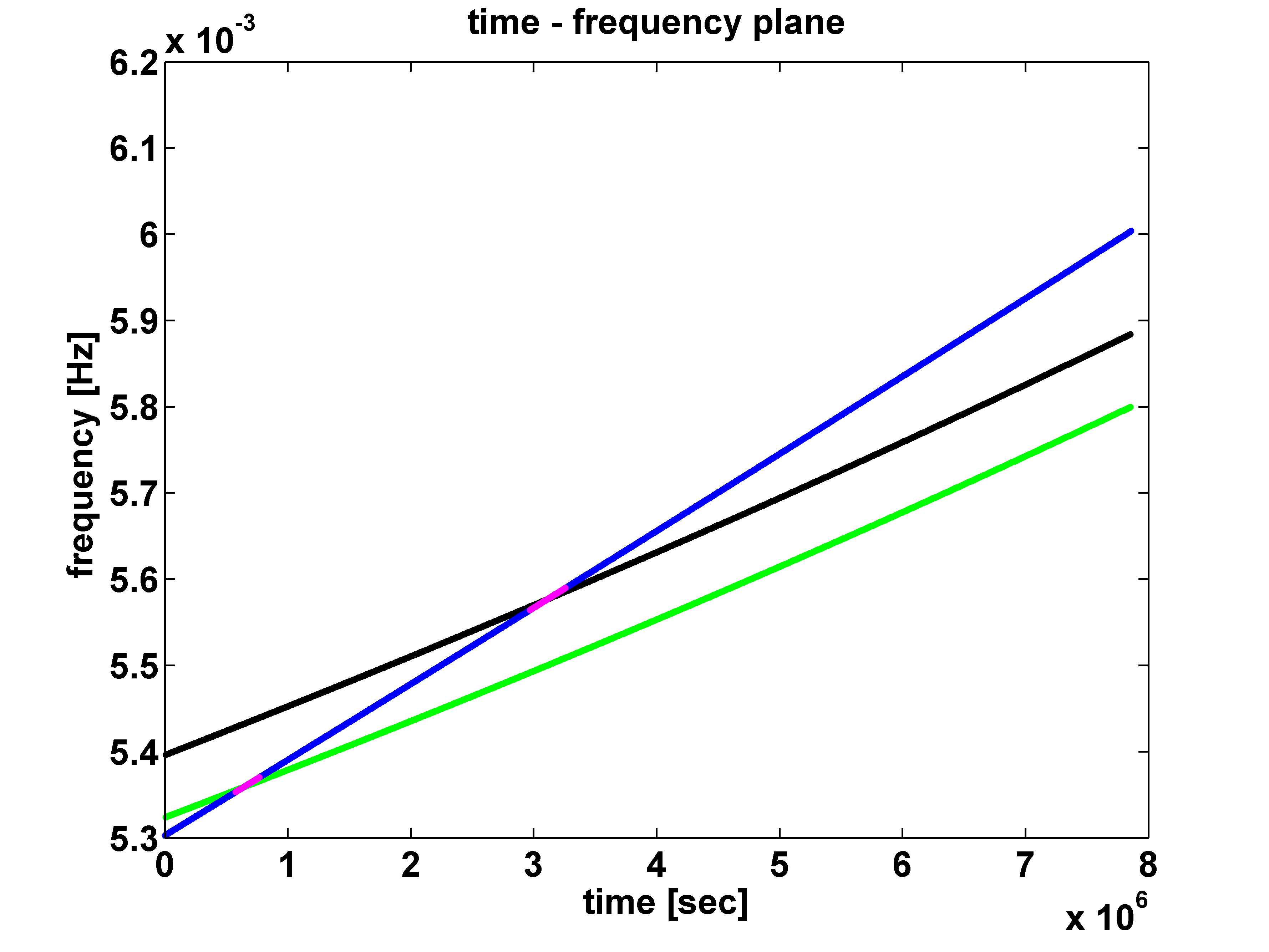}
\caption{ \label{fig:TF_cross} Time-frequency plot of harmonics.
The black and green tracks are two strong harmonics of the EMRI signal
(black being stronger). The blue track corresponds to a harmonic of
PW  that accumulates a significant F-statistic. It intersects the true
harmonics at the pink segments, those correspond to times of increase of
 F-statistic, see Fig.~\ref{fig:unfiltered_CF}, \ref{fig:filtered_CF}.}
\end{figure}

\begin{figure}
\includegraphics[width=0.5\textwidth]{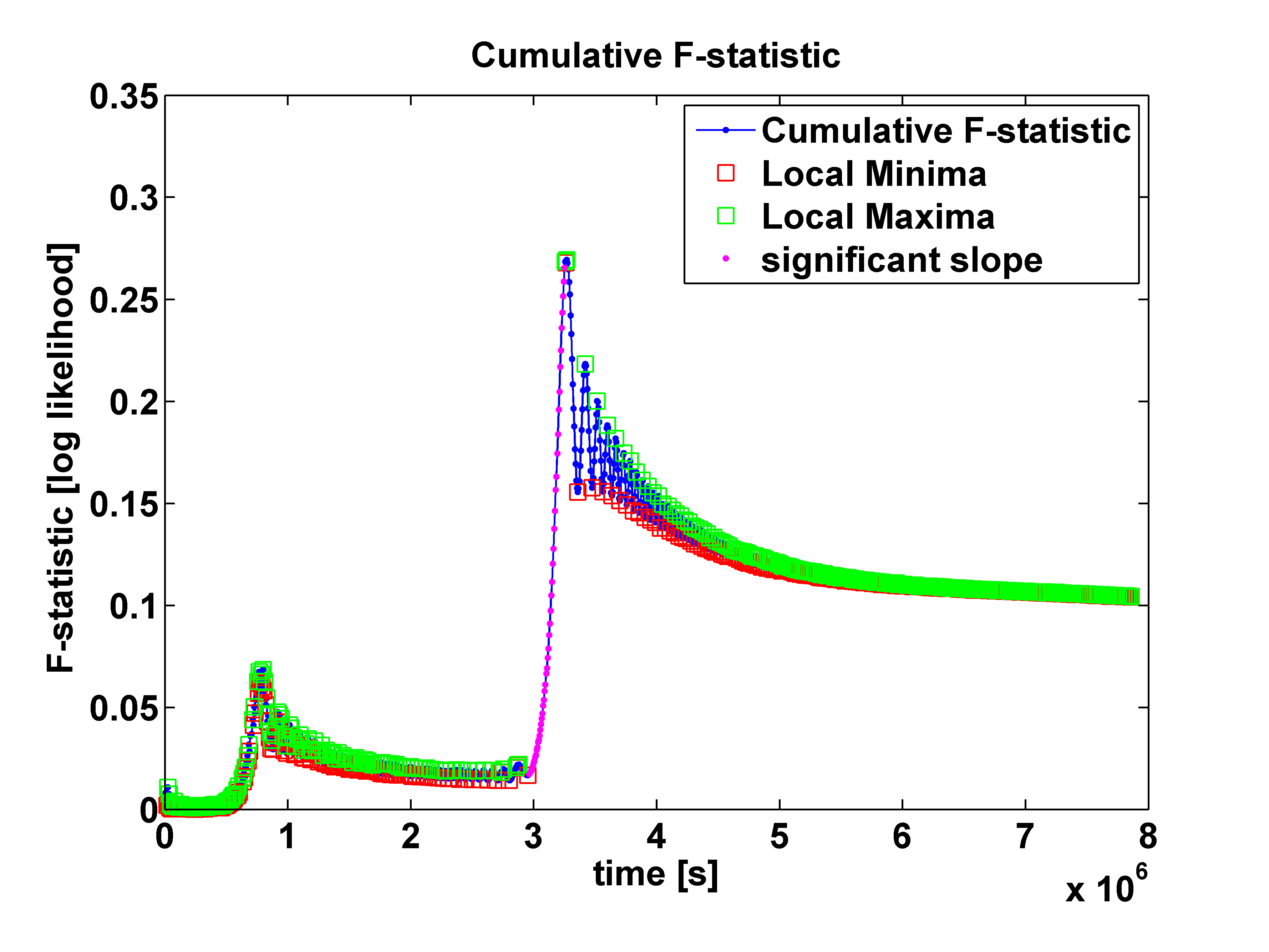}
\caption{ \label{fig:unfiltered_CF} Unfiltered cumulative
F-statistic corresponding to the PW harmonic and data given in
Fig.~\ref{fig:TF_cross}. The F-statistic labeled on the vertical
axis has only relative meaning, since we work with the noiseless data.
The green and red squares mark the extremes of the curve,
thus distinguishing between the increasing and the decreasing
slopes. The large number of the extremes is due to the beating
between the true harmonics and the trial harmonic.}
\end{figure}

  Let us look at one example to understand how we extract the information about the source
from the detected local maxima. We take a particular solution of
MCMC search and for each harmonic of PW we can compute cumulative
F-statistic according to the prescription given in
Section~\ref{S:DAFW}.  We concentrate only on those harmonics which
give significant contribution to the total F-statistic. If the
harmonics of PW match perfectly the harmonics of a signal we should
observe something similar to Fig.~\ref{fig:cumuFsnr50}, however it
is rare when we detect a full harmonic (only sometimes for the
strongest). More frequently, we detect a part of a harmonic
(frequency and derivatives close to true but not exact) or even
several harmonics at different instances of time as shown in
Fig.~\ref{fig:TF_cross}. The black and green curves are two strong
harmonics of a signal (black being stronger), and the blue is a
harmonic of PW.  In the pink regions, our template matches for a
short period of time the frequency of a signal (two distinct
harmonics at two instances). The corresponding cumulative
F-statistic is shown in Fig.~\ref{fig:unfiltered_CF}. There are two
positive jumps in the accumulation of the F-statistic which
correspond to two instances of intersection.  Therefore, we can
conclude that the  positive slope in the cumulative F-statistic (if
it happens over a significant duration) corresponds to the part of
the frequency and time where a harmonic of PW matches (at least
partially) some harmonics of a signal. We collect such events of
matching and display them on the time-frequency plane, resembling
the mosaic  of a true signal.

\begin{figure}
\includegraphics[width=0.5\textwidth]{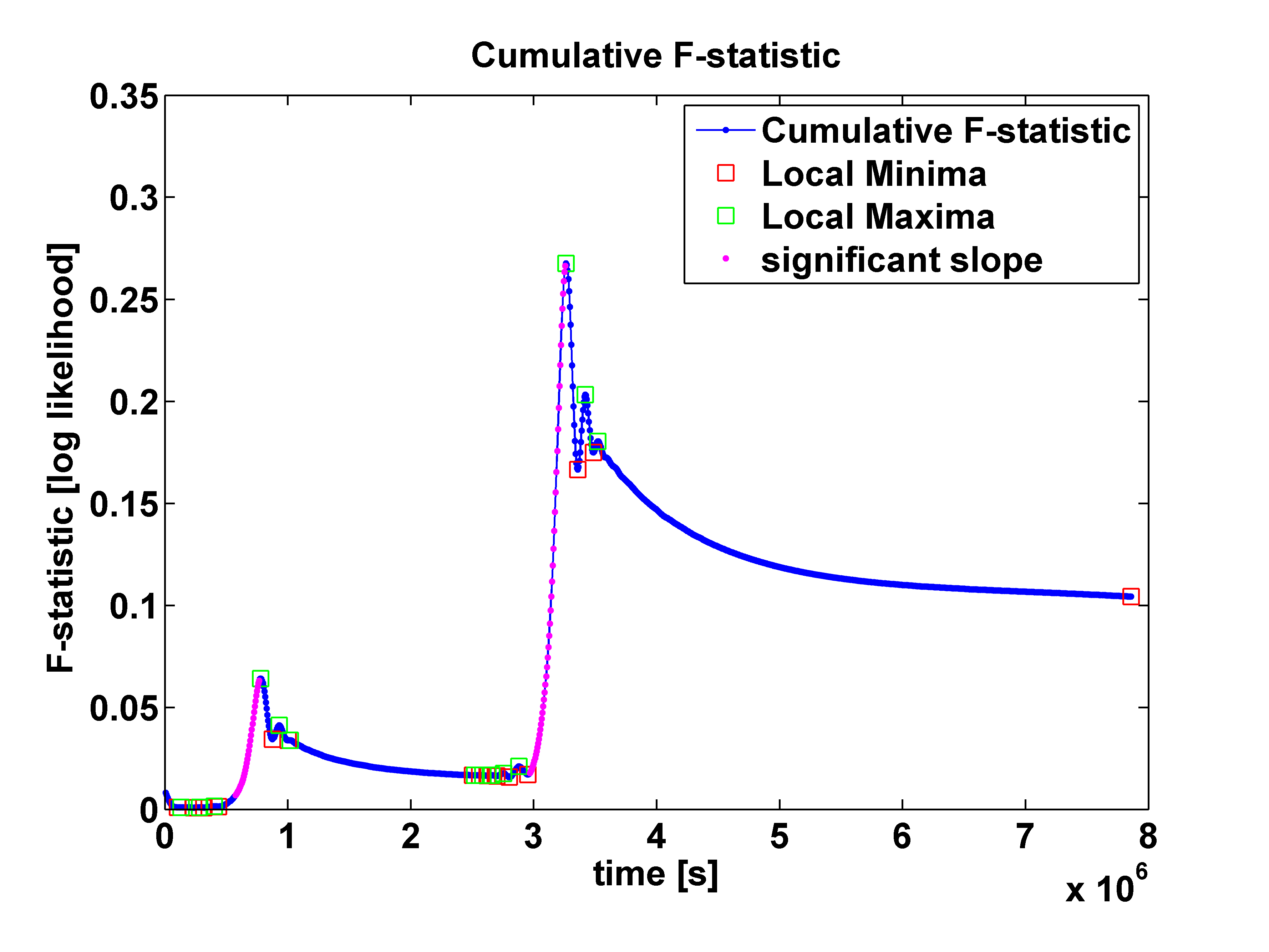}
\caption{ \label{fig:filtered_CF} Filtered cumulative F-statistic
corresponding to the situation depicted in Fig.~\ref{fig:TF_cross}.
 It is similar to Fig.~\ref{fig:unfiltered_CF}, but after
applying the low pass filter to remove the beatings (high frequency
oscillations).}
\end{figure}

 The violent oscillation in Fig.~\ref{fig:unfiltered_CF} is one of the
three types of oscillations on the cumulative F-statistic curve
mentioned in the previous section. In fact, it is the beat note
between the true harmonics and the local maximum. Observe that the
beat notes happen at relatively higher frequency, while the
increasing slopes (where the local maximum matches the frequencies
of the true harmonics) have relative low frequency. Thus, we design
a third-order Butterworth low pass filter to get rid of the beat
notes. After the low-pass filter, the cumulative F-statistic  has
only  few extrema, as shown in Fig.~\ref{fig:filtered_CF}. After
clearing up the cumulative F-statistic, we apply  two criteria for
identifying a significant F-statistic accumulation:  (i) the slope
must be larger than certain threshold; (ii) the accumulation time
must be over longer than certain period. As it is seen by eye tuning
those two parameters should be sufficient to get the right parts of
cumulative F-statistic. In our search we have made the following
choice for those parameters. In the case of noiseless data, we
require the slope to be larger than one-tenth of the largest slope
of the cumulative F-statistic of that trial harmonic, and the
cumulative time (over which we observe steep positive slope) to be
longer than three days.


\begin{figure}
\includegraphics[width=0.5\textwidth]{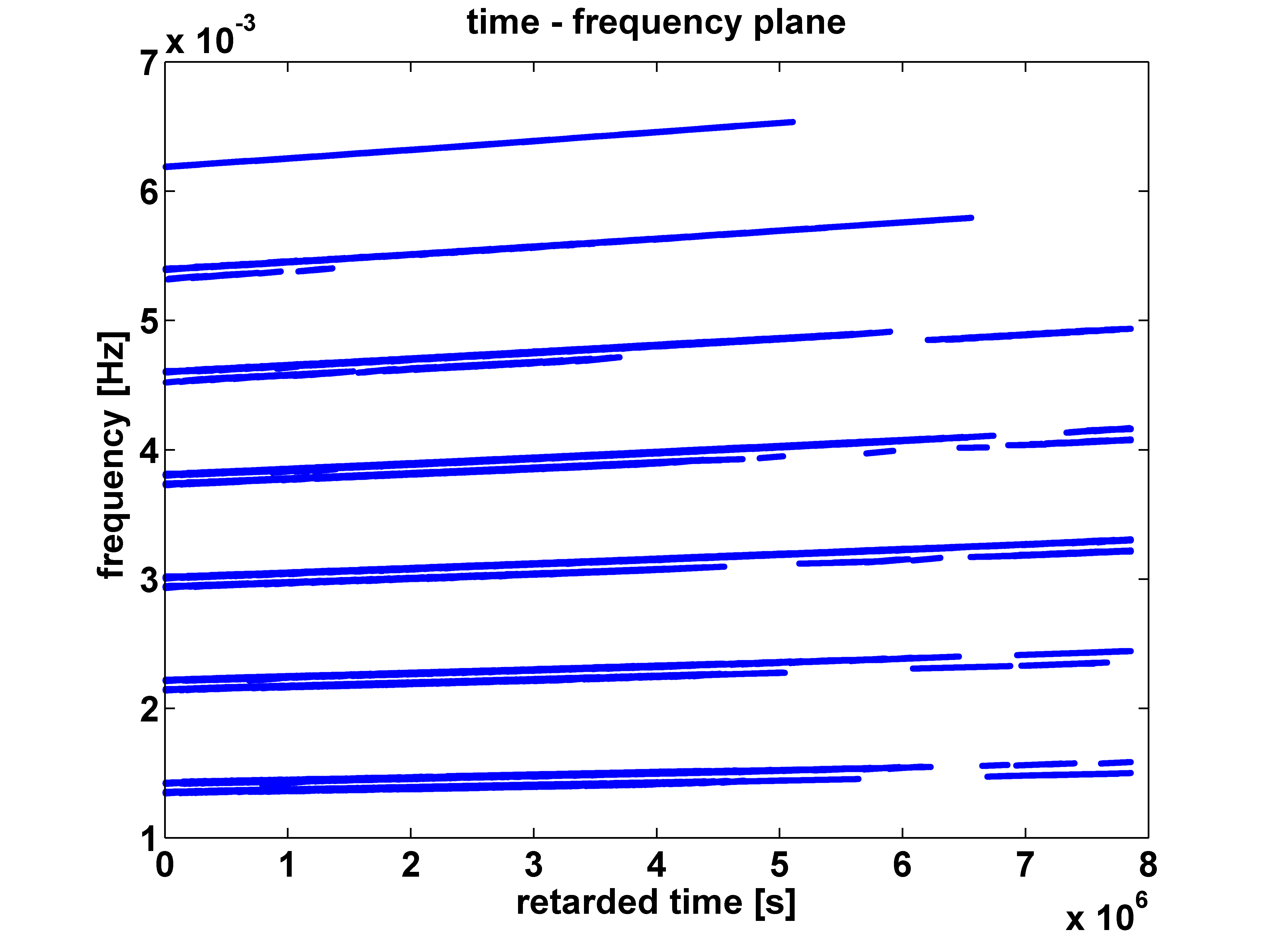}
\caption{ \label{fig:TF_all} Time-frequency plot of all
patches corresponding to strong accumulation of F-statistic.
We can identify parts of frequency tracks of 13 EMRI harmonics.
Each track in this plot has a finite width coming from different
solutions of MCMC search which have different precision of
matching the signal. }
\end{figure}

\begin{figure}
\includegraphics[width=0.5\textwidth]{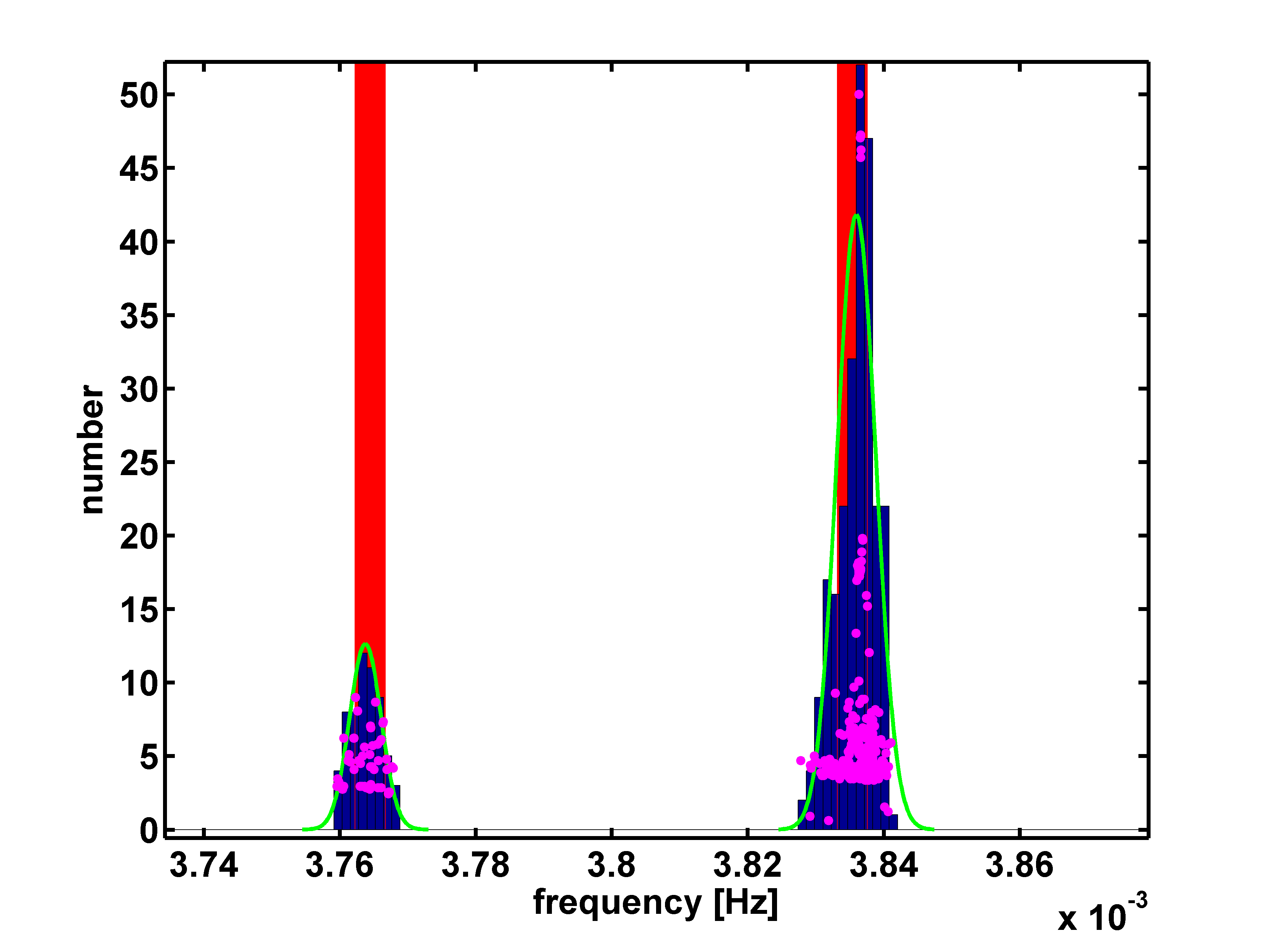}
\caption{ \label{fig:GaussianFit} Zoom at two harmonics at a
specific instance of time. The red stems denote the frequencies of
the true harmonics of a signal, while the blue histogram shows the
detected frequencies at this instant. The green curves display the
Gaussian fit to the frequency data with re-scaled amplitudes. The
vertical axis of pink points indicates the relative time over which
we have observed strong accumulation of F-statistic for each
solution. }
\end{figure}

\begin{figure}
\includegraphics[width=0.5\textwidth]{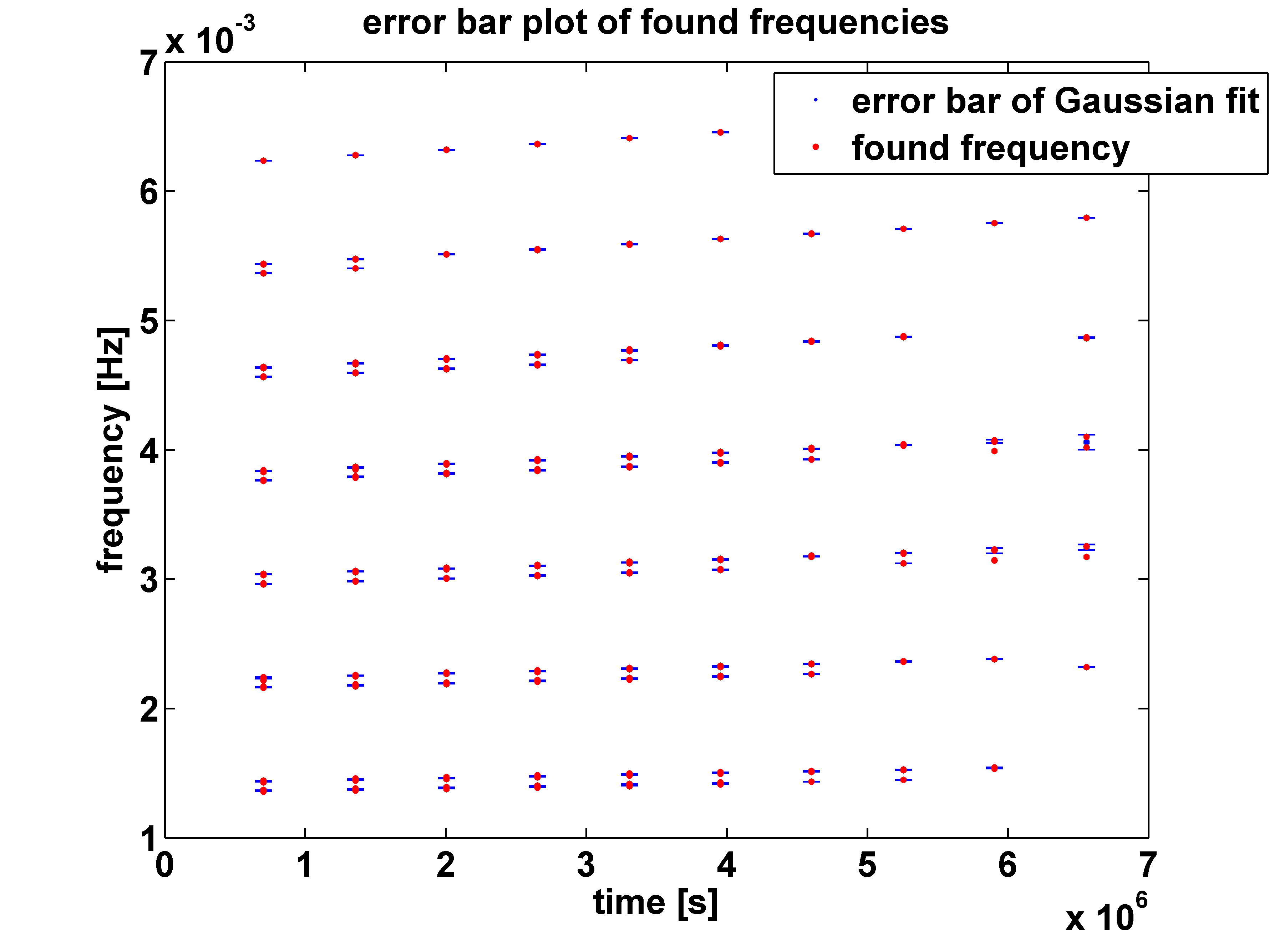}
\caption{ \label{fig:errorbar} Gaussian fit to the detected
frequencies at ten instants. The red points represent the
 mean of a Gaussian fit as shown in Fig.~\ref {fig:GaussianFit}
for each harmonic at ten instants . The blue error bars show the
$1\sigma$ uncertainties of the Gaussian fits.
Note the tiny error bars are along the frequency dimension which indicates that the MCMC search localizes
quite well frequencies of the EMRI's harmonics. }
\end{figure}

 We plot all recovered patches on the time-frequency plane in
  Fig.~\ref{fig:TF_all}, where we can identify by eye
13 strong harmonics. Although the weaker harmonics are lost, the
strong ones retain enough information about the EMRI system
evolution, hence allowing us to recover the physical parameters we
are interested in. Zooming  at a specific harmonic in time and
frequency, one will see that there are many patches from different
results and at each instant we observe a finite spread in the
frequencies for a given harmonic. This is due to various solutions
from MCMC search matched a given harmonic of a signal with different
precision. However, we expect that the distribution of found
frequencies at each instant of time will be centered at the true
frequency of the signal's harmonic. As an example, we show
distribution of found frequencies at a particular instance of time
for two harmonics in Fig.~\ref{fig:GaussianFit}.    In that plot we
show the histogram of detected frequencies at that time in blue and
Gaussian fit as smooth green curves. This is to be compared with
frequencies of two harmonics of a signal at the same time in red. As
mentioned above,  different solutions of MCMC search vary in
precision of matching the signal at different instances, and we can
use accumulation time as a measure of goodness of match of a signal
by a given solution. The relative accumulation time of different
solutions are shown as pink points in Fig.~\ref{fig:GaussianFit}.
First, one can see that Gaussian fit lies on the top of the true
frequency, and second, that the distribution of pink points is
similar to the blue histogram, so either can be taken to
characterize the found harmonics of a signal. Similarly, we can do
at each instance of time for all found tracks in the time-frequency
plane. For the noiseless search we picked uniformly 10 instances and
made a Gaussian fit around each harmonic. We identify the mean of
the Gaussian fit as the most likely frequency of a signal's
harmonics at that instance and we identify the spread (standard
deviation) of a distribution as an error in our evaluation of a
frequency. The result of this clustering is given in
Fig.~\ref{fig:errorbar}.


In the case of data with the detector noise, the basics and the
strategy are roughly the same as in the noiseless case with minor
modifications. In the beginning, we record the local maxima with SNR
greater than $4.5$. Next, we select the significant increasing
slopes of the cumulative F-statistic with three requirements: (i)
the maximum F-statistic along the cumulative F-statistic curve is
larger than $50$, (ii) the minimum slope of the significant
increasing segment is larger than $4\times10^{-6} s^{-1}$, (iii).
the monotonic increasing duration is longer than about a week. Those
conditions are more stringent than for the noiseless case and
eliminated several found weak harmonics of EMRI signal. However, at
the same time they significantly reduce the false events (and that
is what we want). From this selection, we identify 5 strong
harmonics in the noisy case. After that   the procedure is similar
to the noiseless case.

\subsection{Search for physical parameters}

Now we are in a position to recover the physical parameters of the
binary system. First, we need to adopt the model for the orbital
evolution, and here we used the same model as used in the simulation
of the data sets. In the noiseless case the only reason for the
deviation of recovered parameters from the true values is due to
inaccurate identification of the tracks in the time-frequency plane
or due to ambiguity in solving the inverse problem (mapping harmonic
tracks onto the physical parameters, $m/M,  a, e, \iota, p/M$).  We
have performed the search on the time-frequency plane similar in
spirit to \cite{Gair:2008ec}. We have used weighted chi-square test

$$
\chi^2 = \sum\min_{f_{lmn}}
\left(\frac{f_{lmn}-f_\mathrm{mean}}{\sigma_f}\right)^2
$$

between signal tracks (for different parameters) and recovered
tracks (Fig.~\ref{fig:errorbar}).
    We have used particle swarm optimization (PSO) and genetic algorithm (GA) as two independent search methods
to test the robustness of our result. We start with describing the PSO method, and then give brief overview of GA.

Particle swarm optimization (PSO) is a stochastic optimization
method introduced by Kennedy and Eberhardt in 1995~\cite{Kennedy95}.
In gravitational wave data analysis, PSO was first applied to a
binary inspiral signal~\cite{Wang:2010jma}. In this section, we
briefly describe the algorithm, while further details can be found
in the
references~\cite{Kennedy95,Wang:2010jma}.\\
The goal of PSO is to find the global minimum/maximum (here we minimize the chi-square test)
 of a parameterized functional $\kappa(\bm{\theta})$ and the corresponding parameter set
$\bm{\theta}_*$, where $\bm{\theta}$ stands for an arbitrary
parameter set in $\mathbb{R}^n$. The idea is to evaluate
$\kappa(\bm{\theta}_i)$ simultaneously at different parameter sets
$\bm{\theta}_i,\,i=1,2,...$, treating them as particles in the
parameter space, and evolve them according to certain dynamics until
the stable solution  is reached. Let us denote the $i$-th particle out of a
swarm of $N_p$ particles during $k$-th iteration in the search by
$\bm{\theta}_i[k]$. Its position in the parameter space in the next
iteration is determined by its velocity in the current iteration
$\bm{v}_i[k]$,
\begin{eqnarray}
\bm{\theta}_i[k+1]=\bm{\theta}_i[k]+\bm{v}_i[k].
\end{eqnarray}
Usually, the particles start with randomly chosen positions
$\bm{\theta}_i[1]$ and velocities $\bm{v}_i[1]$. Up to $k$-th
iteration, we denote the $i$-th particle's best location by
$\bm{\theta}^p_i[k]$, in the sense that
\begin{eqnarray}
\kappa(\bm{\theta}^p_i[k])=\min_{j\leq k} \kappa(\bm{\theta}_i[j]).
\end{eqnarray}
The global best location $\bm{\theta}^g_i[k]$ up to the $k$-th
iteration is defined by
\begin{eqnarray}
\kappa(\bm{\theta}^g[k])=\min_{i} \kappa(\bm{\theta}^p_i[k]).
\end{eqnarray}
Note that particle best locations and the global best location are
the best parameters respectively found by the individual particles
and the whole swarm in the entirely history of the search up to the
$k$-th iteration. They are updated only when a better parameter set
is found. These best locations contain a lot of information about
the functional $\kappa(\bm{\theta})$, so they are used to guide the
particle's motion in the future. Explicitly, the velocities are
updated with the following equation
\begin{eqnarray}
\bm{v}_i[k+1] = w\bm{v}_i[k] +
c_1\chi_1(\bm{\theta}^p_i[k]-\bm{\theta}_i[k])+  \nonumber\\
c_2\chi_2(\bm{\theta}^g[k]-\bm{\theta}_i[k]),
\end{eqnarray}
where $w$ is called the {\em inertia weight}, $c_1,\,c_2$ are called
the {\em acceleration constants} (we take them to be the same as in
\cite{Wang:2010jma}) and $\chi_1,\,\chi_2$ are random numbers drawn
from $\mathcal{U}(0,1)$. We run PSO search several times until the
return result is confirmed by several
searches. \\

The second search method is called Genetic Algorithm (GA) and there we evolve
a number of parameter sets (points in  the parameter space $\mathbb{R}^n$).
Each parameter set   $\bm{\theta}_i$ is called an organism, individual
parameters are called the genes of this organism and the set of organism
at $k$-th search iteration step is called $k$-th generation.
We evolve generations according to the prescribed rules called
``parents selection'', ``breading'' and ``mutation''. The main idea of this
optimization technique is to evolve colony of organisms toward the better
fitness  (which could be likelihood ratio or, in our case, chi-square value)
like in Darwin's theory of natural selection. The strong organisms (with better
fitness) participate more often in breading and therefore drag the colony
toward the better values (lower) of chi-square. Mutation brings element of randomness
   in the search and occasional ``positive'' mutations help to avoid trapping
around local minimum. For use of GA in GW data analysis we refer to
\cite{Crowder:2006wh, Petiteau:2010zu} and references therein.

 Let us give few more details specific to the implementation used in here.
We use $\chi^2$ value as a measure of fitness for each organism
(smaller value is better). In each generation we use the roulette
method with the selection probability proportional to the fitness of
each organism. For breeding we have used the one random point
crossover  rule. The probability mutation rate is monotonically
decreasing function of the generation number: we have started with
high probability of mutation to explore a large part of the
parameter space and decrease it gradually as organisms converge to a
particular part of the parameter space.  We  have used ``children''
and  ``parents'' sorted in the fitness to make a new generation: we
use 50\% of the best organisms. We automatically achieve the
``elitism'' in a way that the best $\chi^2$ value is never
increasing from one generation to the next.

We use the multi-step method to accelerate the search. In each step we evolve the colony for 500 generations as described above, but
each new step uses the last generation of the previous step as the initial state. We have started evolution in the first step with
completely random distribution of the organisms.  The evolution of the colony at each step finishes with a very small mutation probability
and with organisms confined to a quite small volume of the parameter space. The consequent search steps ensure that the found solution
is a robust solution with respect to increase of the mutation probability which disperses organisms forcing them to explore the parameter
space for presence of a solution with better fitness. This helps to avoid being trapped in the local minima. The termination condition is the stability
of the best solution over several steps of the search.
\\

We have applied both those methods to fit the found tracks on the
time frequency plane with the harmonics of EMRI signal. The search
is done in 5 dimensional parameter space with quite broad priors on
$(e, p/M, \iota, a, \mu= m/M)$, those are the eccentricity, the
semi-latus rectum, the orbital inclination angle at the moment of
beginning of observation, the spin of the MBH, and, the mass ratio
between the stellar BH and the MBH.   The total mass is not present
here, we have kept it fixed to
 $M = 10^6 M_{\odot}$.  For a given set of parameters, our search
algorithm computes three fundamental orbital frequencies as
functions of time, then a weighted chi-square goodness of fit test
is preformed on harmonics of the signal. We use the means and
standard deviations from the Gaussian fit as found point and its
error in the time-frequency plane.    The best fit corresponds to
the lowest value of $\chi^2$. We have used harmonics of the signal,
which are expected to be strong over the large part of the parameter
space, and have found this ``harmonic table'' by intensive monte
carlo with NK models generated in the frequency domain. The index
table has been truncated by choosing harmonics contributing (in
total) 90\% of the overlap with a total signal~\footnote{The total
signal here to be a NK waveform with a large number of harmonics. We
still truncate the number of harmonics used to build the signal: we
stop if the inclusion of the next harmonic does not change overlap
with the already built signal by more than 0.1\%.}.

The recovered parameters are given in
the table~\ref{T:result}.

\begin{table*}
\centering

\caption {\label{T:result} Recovered parameters of EMRi against
actual parameters used in simulated data sets. }
\begin{tabular}{@{}c|ccccc@{}}
\hline
 description & $ e(t_0)$ & $p(t_0)$ & $\iota(t_0)$ & a & $\mu$ \\
\hline
True  parameters  &  0.4 & 8.0 & 0.349  &  0.9 & $10^{-5}$ \\
Recovered parameters (with noise)  &  0.395 & 8.029 & 0.342 & 0.891 & $9.79 \times 10^{-6}$ \\
Recovered parameters (no noise) & 0.402 & 7.991& 0.360  & 0.901 & $1.002 \times 10^{-5}$   \\
\hline
\end{tabular}
\end{table*}

\section{Summary}
\label{S:Summary}

 In this paper we have introduced the phenomenological family of waveforms (PW) for detecting EMRI signals
in the data from the LISA-like observatory. The template is
constructed out of independent (over the time interval we have
applied our analysis) harmonics of slowly evolving three orbital
frequencies. We have neglected the amplitude evolution and presented
the phase as a Taylor series up to the third derivative of
frequency.  Our analysis was restricted  to the case of
monotonically increasing frequencies. This condition will break only
close to the plunge. The number of harmonics and range of indices
were taken from the analysis of dominant harmonics of our model
signal, though we have found at the end that the search only weakly
depends on the number of used harmonics (only through the
accumulated total SNR, which should be sufficient to claim
detection).

Constructed  phenomenological templates allows us to search for EMRI
signals in a model independent way.  This way we avoid complexity of
accurate modeling the orbital evolution and gravitational waveform
during the search. In addition PW cover also all possible small
deviations of the background spacetime from the Kerr solution which
would influence the signal's phase and could lead even to loss of
the signal if the template assumes pure Kerr background geometry.

We have used MCMC based search to find a large number of local
maxima of the likelihood surface. We were not that lucky to find the
global maximum. We have analyzed the found solutions by means of
cumulative F-statistic over the time and identified the patches of
the signal which were match by templates. As a result, we have
constructed a time-frequency map of (parts of) the signal's
harmonics. Each track could be  characterized by the best guess and
the error bar at each instance of time (by fitting Gaussian profile
to found frequencies around at that time each track).  The next step
is to assume a model for the binary orbital evolution, and check if
the found time-frequency picture corresponds to the strongest
harmonics of a signal.  In other words, we want to find the physical
parameters of the binary system which strong GW harmonics could
leave the found imprint. We do that by conducting a search using
particle swarm optimization techniques and, independently, genetic
algorithm. We have used weighted chi-square goodness of fit test to
choose the best matching harmonics of the signal. We have assumed
the same model as was used in the simulated data, and the recovered
parameters are within 2\% of the true values.

We want to make few final remarks. (i) The found time-frequency
tracks of the GW signal from EMRI did not assume any particular
model. The mapping of these tracks to the physical parameters could
be done in post processing using several models. We have chosen on
purpose rather short (3 month) duration of the data. The search
procedure could be repeated for each three months and then one can
check consistency of a given model or further improve accuracy in
the recovered parameters (if our model gives consistent parameters
of the system across different data segments).  This could be a
powerful method to search deviations from ``Kerness''.  (ii) In the
mapping of the time-frequency tracks to the physical parameters of
the binary, we have only weakly used information about the strength
of each track/harmonic. We have found that the information stored in
the frequency evolution is sufficient to recover parameters of
EMRI. However, additional information about the strength
of the recovered harmonics and harmonics of the modeled GW signal
could give us additional confidence in the result and/or distinguish
between several solution, if ambiguity happens. (iii) Mapping from the
found time-frequency tracks onto the physical parameters might turn
out to be the most computationally intensive task. However, one might
use the information about the strength and a number of found
harmonics to restrict a volume of the searched  parameter space. In
addition, to perform mapping we require mainly the computation of the
orbital evolution, not the full waveform. However, it is then
important to know which harmonics are the strongest for a given
parameter set.   (iv) In the future work we intend to include the
sky location and the MBH mass into the search and investigate the
possibility to differentiate between different models of EMRIs based
on the results of MCMC search with PW (as discussed in (i)).

\section{Acknowledgement}

The authors would like to thank S. Drasco and J. Gair for useful discussions.
The work of S. B. and Y. W. was partially supported by DFG Grant No. SFB/TR 7 Gravitational Wave
Astronomy and DLR (Deutsches Zentrum fur Luft- und Raumfahrt). Y. W.
also would like to thank the German Research Foundation for funding
the Cluster of Excellence QUEST-Center for Quantum Engineering and
Space-Time Research. Y. S. was supported by MPG within the IMPRS
program.

\end{document}